\documentclass[aps,twocolumn,preprintnumbers,nofootinbib,superscriptaddress]{revtex4-1}
\usepackage{times}
\usepackage{amsmath}
\usepackage{amssymb}
\usepackage{graphicx}
\usepackage{subfigure,accents}
\usepackage{booktabs}
\usepackage{rotating}
\usepackage{longtable}
\usepackage{bm}
\usepackage[pagebackref=false,colorlinks=,linkcolor=blue,citecolor=blue]{hyperref}
\usepackage{epstopdf}
 \pdfoutput=1
 \usepackage[utf8]{inputenc}
\usepackage{cases}
\usepackage{amsmath,float}
\usepackage{amssymb}
\usepackage{amsfonts}
\usepackage{amssymb}
\usepackage{dcolumn}
\usepackage{bm}
\usepackage{bbm}
\usepackage{graphicx}
\usepackage{xcolor}
\usepackage{array}
\usepackage{subfigure}
\usepackage{hyperref}
\usepackage{wasysym}

\newcommand{\be}{\begin{equation}}
\newcommand{\ee}{\end{equation}}
\newcommand{\ba}{\begin{eqnarray}}
\newcommand{\ea}{\end{eqnarray}}

\newcommand{\gsim}{\mathrel{\hbox{\rlap{\lower.55ex \hbox {$\sim$}}
                   \kern-.3em \raise.4ex \hbox{$>$}}}}
\newcommand{\lsim}{\mathrel{\hbox{\rlap{\lower.55ex \hbox {$\sim$}}
                   \kern-.3em \raise.4ex \hbox{$<$}}}}

\hypersetup{colorlinks=true,
            breaklinks=true,
            pdfstartview=Fit,
            linkcolor=blue,
            citecolor=blue,
            urlcolor=blue}
            
 \usepackage{graphicx}

\usepackage{amsmath}

\newcommand{\bw}{\begin{widetext}}
\newcommand{\ew}{\end{widetext}}

\newcommand{\diag}{\ensuremath{ \text{diag} }}

\def\ber{\begin{eqnarray}}
\def\eer{\end{eqnarray}}
\def\beq{\begin{equation}}
\def\eeq{\end{equation}}

\begin{document}

\title{Regular black holes in three dimensions and the zero point length
}
\author{Kimet Jusufi}
\email{kimet.jusufi@unite.edu.mk}
\affiliation{Physics Department, State University of Tetovo, Ilinden Street nn, 1200, Tetovo, North Macedonia}

\begin{abstract}
    In this paper, by means of regularisation procedure via $r\to \sqrt{r^2+l_0^2}$ (where $l_0$ can play the role of zero point length), we first modify the gravitational and electromagnetic potentials in two dimensions and then we solve the Einstein field equations to end up with an exact and regular black hole solution in three dimensions with a negative cosmological constant. We show that, the black hole solution is asymptotically AdS, non-singular at the origin and, under specific conditions, it has a flat de Sitter core at the origin. As a special case, we obtain the charged Banados-Teitelboim-Zanelli (BTZ) solution. Finally, using a dimensional continuation and the  NJ algorithm, we end up with a legitimate rotating black hole solution in three dimensions.  
\end{abstract}
\maketitle

\section{Introduction}
Black holes solutions was believed to cannot exist in three spacetime dimensions. This line of thought was based on the no local gravitational attraction and therefore no mechanism to produce black holes. It came as a surprise, when Banados, Teitelboim, and Zanelli (BTZ) \cite{BTZ1,BTZ2} reported a vacuum black hole solution in three spacetime dimensions with Anti-de Sitter (AdS3) space. The BTZ black hole solution was an exact solution of Einstein field equations, besides the mass parameter, it was shown that such a black holes can have electric charge and can rotate. Another aspect of the BTZ black hole is presence of singularity at the origin. The BTZ solution gained a lot of interest in the community and the interested reader is referred to some interesting works in this direction \cite{3,4,5,6,7,8} and references therein. The solutions in  AdS are of interest due to the AdS/CFT correspondence proposed by Maldacena \cite{M}. This correspondence was studied for the case of BTZ spacetime in AdS3 \cite{9}. 

In this work, we are interested to construct regular black holes in three dimensions using a regularisation procedure via $r\to \sqrt{r^2+l_0^2}$, where $l_0$ can play the role of zero point length. Such a regularisation procedure is closely related to ideas from T-duality. The T-duality theory is an equivalence between two different string theories in two contexts. Precisely speaking, by T-duality theory, with the winding modes and identification of momentum, one can relate the geometries having large and small compact directions. Recently, a four dimensional quantum corrected black holes in T-duality was found \cite{td1}. Such a black hole was shown to be regular and coincides with the Bardeen solution. In addition, for an exact black hole solution with charge in T-duality see \cite{td2}. In the present work, we would like to extend this idea to study regular black holes in three dimensions.

The paper is structured as follows. In Section II, we modify the gravitational and electromagnetic potentials and we solve the Einstein field equations in three dimensions. In Section III, we use dimensional continuation and the NJ algorithm in order to generate a rotating black hole solutions. In Section IV, we comment on our findings. 

\section{Regular and charged black holes in three dimensions}
It was recently shown (see for details \cite{td1,td2}) that by using ideas from T-duality  one can modify the standard Newtonian potential [and similarly the electric potential] as follows $\Phi(r)\to - M/\sqrt{r^2+l_0^2}$. The gravitational potential in two dimensions on the other hand is given by $\Phi \sim M \ln(r)$, with $k$ being some constant and $M$ is mass per unit surface. We now impose the regularization factor $r \to \sqrt{r^2+l_0^2}$, to the potential which results with
\begin{equation}
    \Phi(r)=k M \ln\left( \sqrt{r^2+l_0^2}\right).
\end{equation}
Now solving the Poisson's equation in polar coordinates we can obtain the energy density
\begin{equation}
   \nabla^2 \Phi(r)=2 \pi \rho,
\end{equation}
yielding 
\begin{equation}
    \rho(r)=\frac{k\,M\,l_0^2}{\pi (r^2+l_0^2)^2},
\end{equation}
meaning that when $l_0\to 0$, it reduces to zero, i.e., point mass. In other words, due to the zero point length, we get a smeared matter distribution due to quantum gravity effects described by the quantum modified energy momentum   tensor
${(T^\mu}_\nu)^{corr} = \diag\left(-\rho,p_r,p_{\phi}\right)$. Here, $p_r$ and $p_{\phi}$ are the radial and transverse pressures, respectively. For the mass function of the
black hole we get
\begin{eqnarray}
m(r)=-2 \pi \lim_{r \to \infty} \int_{0}^{r} r'  [T^{0}]_0 dr'=\frac{k M r^2}{l_0^2+r^2}.
\end{eqnarray}
If we further set $k=1$ and $l^2_0\to 2ML$, this mass profile has a similar form to the Hayward model used in four dimensions although not exactly the same as here it is used in context of three dimensional case. Furthermore, such a  profile has been conjectured recently in \cite{8}. It is thus  very interesting to see that we obtain this profile using the regularisation procedure. We are now interested to find a three dimensional black hole solution derived from the Einstein-Maxwell action and quantum effects using the action 
\begin{equation}
    S=\int  \sqrt{-g} \left[ \frac{R}{2 \kappa }+2\Lambda -\frac{1}{4}F^{\mu \nu}F_{\mu \nu}\right]dx^2+S_{corr}
\end{equation}
where the cosmological constant $\Lambda=1/l^2>0$ in our notation.  The line element of the black hole in cylindrical coordinates reads
\begin{equation}
    ds^2=-f(r)dt^2+\frac{dr^2}{f(r)}+r^2 d\phi^2.
\end{equation}

Solving for the field equations one can find the Einstein-Maxwell equations 
\begin{equation}
    G_{\mu \nu}-\Lambda g_{\mu \nu}=\kappa \left(T_{\mu \nu}^{corr}+T_{\mu \nu}^{em}\right),
\end{equation}
\begin{eqnarray}
\nabla_{\mu}F^{\mu \nu}=4 \pi J^{\nu}
\end{eqnarray}
with the energy-tensor for the electromagnetic field given by 
\begin{equation}
 T_{\mu \nu}^{em}=\frac{1}{4 \pi}\left(F_{\mu \sigma}{F_{\nu}}^{\sigma}    -\frac{1}{4}g_{\mu \nu}F_{\rho \sigma}F^{\rho \sigma}\right).
\end{equation}
In the spirit of Eq. (1) we regularise now the electromagnetic potential as follows
\begin{equation}
A_\mu=(-Q \ln(\sqrt{r^2+l_0^2}),0,0).
\end{equation}

Using the spacetime metric (6) we get the only non-vanishing components of the Faraday tensor 
\begin{equation}
F_{tr}=-F_{rt}=\frac{Q r}{r^2+l_0^2},
\end{equation}
along with the scalar quantity
\begin{equation}
F_{\mu \nu}F^{\mu \nu}=-\frac{2Q^2 r^2}{(r^2+l_0^2)^2}.
\end{equation}

For the energy-momentum tensor of the electromagnetic field we find the components:
\begin{eqnarray}
    {\rho}^{em}(r)&=&-{p_r}^{em}(r)=\frac{Q^2 r^2}{8\pi\left(r^2+l_0^2\right)^2},\\
    {p_{\phi}}^{em}(r)&=&\frac{Q^2 r^2}{8 \pi\left(r^2+l_0^2\right)^2}.
\end{eqnarray}

\begin{figure}[]
    \centering
         \includegraphics[scale=.41]{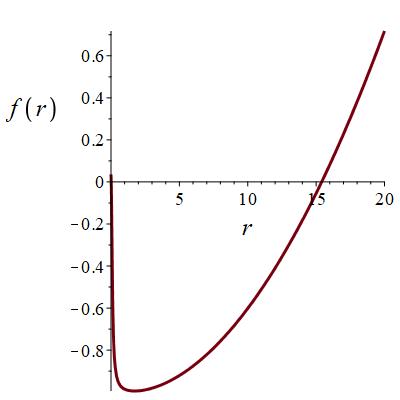}
       \caption{The plot shows the location of horizons using $f(r)$. We used $M=1$, $l_0=0.1$, $Q=0.1$ and $l=15$. The outer/inner horizons are located at $r_+=15.411$ and $r_{-}=0.023$, respectively. }\label{f1}
\end{figure}

From the Einstein field equations and energy momentum tensor we obtain the $t-t$ component given by 
\begin{eqnarray}
\frac{f'(r)}{2 r}+\frac{Q^2 r^2}{\left(r^2+l_0^2\right)^2}+\frac{M\,l_0^2}{ (r^2+l_0^2)^2}-\frac{1}{l^2}=0,
\end{eqnarray}
where for simplicity we have set $k=1/8$. One can then easily obtain the following exact solution
\begin{equation}
    f(r)=\mathcal{C}+\frac{r^2}{l^2}+\frac{M l_0^2}{r^2+l_0^2}-\frac{Q^2l_0^2}{r^2+l_0^2}- Q^2\ln(r^2+l_0^2),
\end{equation}
where the integration constant can be fixed after we take the limit $l_0\to 0$, which should yield the standard result \cite{BTZ1,BTZ2}
\begin{equation}
    f(r)=\frac{r^2}{l^2}-M-2 Q^2\ln(r),
\end{equation} 
hence, we can get $\mathcal{C}=-M$. Thus, the general result of our solution is given by
\begin{equation}
    f(r)=\frac{r^2}{l^2}-\frac{M r^2}{r^2+l_0^2}-\frac{ Q^2l_0^2}{r^2+l_0^2}- Q^2\ln(r^2+l_0^2).
\end{equation}
To the best of our knowledge this metric is new and not obtained yet in the literature. The black hole solution is regular for any $r$, including the limit $r \to 0$.  To see this, let us show here the Ricci scalar 
\begin{eqnarray}
 \lim_{r \to 0}   R_{\mu \nu} R^{\mu \nu}=\frac{6 M}{l_0^2}-\frac{6}{l^2},
\end{eqnarray}
and the Kretschmann scalar which reads
\begin{eqnarray}\notag
 \lim_{r \to 0}   R_{\mu \nu \sigma \rho} R^{\mu \nu \sigma \rho}&=&\frac{12M^2}{l_0^4}-\frac{24 M}{l_0^2 l^2}+\frac{12}{l^4},
\end{eqnarray} 
provided $l_0>0$. The solution for large $r$ are asymptotically AdS. Horizons of the metric are roots of the lapse function $f(r)$. In general there is an outer and inner horizon, see Fig. 1. The extreme case, i.e., when the real roots gives a single horizon is found the condition $f'(r)=0$. Let us point out here that, we can consider the case $r<<l_0$, and set the charge to zero for simplicity, i.e., $Q=0$, then from Eq. (16) we get 
\begin{eqnarray}
f(r)=M-\mathcal{C}-\left(\frac{M}{l_0^2} - \frac{1}{l^2} \right)r^2+...
\end{eqnarray}
this means that in specific conditions the quantum gravity effects can produce a de Sitter core at the origin 
\begin{eqnarray}
f(r)=1-\Lambda_{eff}r^2+...
\end{eqnarray}
where we have set $M-\mathcal{C}=1$ and $\Lambda_{eff}=M/l_0^2-1/l^2>0$, which $M>l_0^2/l^2$. A similar result was shown in a recent work \cite{8}.

\section{Generating rotating solutions}
In this section we like to generate a rotating black hole solution in three dimensions. In doing so, we shall apply the complex coordinate transformation scheme used in Ref. \cite{Kim:1998iw}, however here we shall generalize it for any function $f(r)$. The main idea behind this derivation is to assume that our solution in three dimensions is a slice of a static and spherically symmetric of a four geometry given by
\begin{equation}
ds^2_{4} = -f(r)dt^2 + f(r)^{-1}dr^2 + h(r) (d\theta^2 +
\sin^2 \theta d\phi^2)
\end{equation} 
with $f(r)$ given by Eq. (18), and $h(r)=r^2$. This 
``dimensional continuation" basically allows us to introduce the null tetrad system of vectors and therefore to make use of the Newman's complex coordinate transformation method. As was nicely explained in \cite{Kim:1998iw}, ans we note here also that we do not demand that this four dimensional geometry to be a solution of Einstein equations, but we expect its slice [for $\theta = \pi/2$] to be a solution of Einstein equation in three dimensions. Let us now apply the NJ method, however first we have to rewrite metric (22) in Eddington-Finkelstein-type retarded 
null coordinates $(u, r, \phi)$, given by
\begin{eqnarray}
ds^2_{4} = -f(r)du^2 - 2du dr + h(r) (d\theta^2 +
\sin^2 \theta d\phi^2).
\end{eqnarray}
On the other hand, this metric can be decomposed in terms of null tetrads as follows
	\begin{eqnarray}
		g^{\mu{\nu}}=-l^{\mu}n^{\nu}-l^{\nu}n^{\mu}+m^{\mu}\overline{m}^{\nu}+m^{\nu}\overline{m}^{\mu},
	\end{eqnarray}
	along with the null vectors defined as
	\begin{eqnarray}
		l^{\mu}&=&\delta^{\mu}_{r},\\
		n^{\mu}&=& \delta^{\mu}_{u}-\frac{1}{2} f(r)\delta^{\mu}_{r},\\
		m^\mu&=&\frac{1}{\sqrt{2}\,r}\left(\delta^{\mu}_{\theta}+\frac{\dot{\iota}}{\sin\theta}\delta^{\mu}_{\phi}\right),\\
		\overline{m}^\mu&=&\frac{1}{\sqrt{2}\,r}\left(\delta^{\mu}_{\theta}-\frac{\dot{\iota}}{\sin\theta}\delta^{\mu}_{\phi}\right).
	\end{eqnarray}
	In the above notation $\bar{m}^\mu$ is the complex conjugate of $m^\mu$. In particular, these vectors further satisfy the conditions for normalization, orthogonality and isotropy as
	\begin{eqnarray}
		l^{\mu}l_{\mu}=n^{\mu}n_{\mu}=m^{\mu}m_{\mu}=\bar{m}^{\mu}\bar{m}_{\mu}=0,\\
		l^{\mu}m_{\mu}=l^{\mu}\bar{m}_{\mu}=n^{\mu}m_{\mu}=n^{\mu}\bar{m}_{\mu}=0,\\
		-l^{\mu}n_{\mu}=m^{\mu}\bar{m}_{\mu}=1.
	\end{eqnarray}
	Following the NJ prescription and its modified version \cite{Azreg-Ainou:2014pra} we write,
	\begin{equation}
		{x'}^{\mu} = x^{\mu} + ia (\delta_r^{\mu} - \delta_u^{\mu})
		\cos\theta \rightarrow \\ \left\{\begin{array}{ll}
			u' = u - ia\cos\theta, \\
			r' = r + ia\cos\theta, \\
			\theta' = \theta, \\
			\phi' = \phi. \end{array}\right.
	\end{equation}
	in which $a$ stands for the rotation parameter. Next, let the null tetrad vectors $Z^a=(l^a,n^a,m^a,\bar{m}^a)$ undergo a
	transformation given by $Z^\mu = ({\partial x^\mu}/{\partial {x^\prime}^\nu}) {Z^\prime}^\nu $, following
	\begin{eqnarray}
		l'^{\mu}&=&\delta^{\mu}_{r},\\
		n'^{\mu}&=&\delta^{\mu}_{u}-\frac{1}{2}F\delta^{\mu}_{r},\\\label{e11}
		m'^{\mu}&=&\frac{1}{\sqrt{2\,\Sigma}}\left[(\delta^{\mu}_{u}-\delta^{\mu}_{r})\dot{\iota}{a}\sin\theta+\delta^{\mu}_{\theta}+\frac{\dot{\iota}}{\sin\theta}\delta^{\mu}_{\phi}\right],\\
		\overline{m}'^{\mu}&=&\frac{1}{\sqrt{2\,\Sigma}}\left[(\delta^{\mu}_{u}-\delta^{\mu}_{r})\dot{\iota}{a}\sin\theta+\delta^{\mu}_{\theta}+\frac{\dot{\iota}}{\sin\theta}\delta^{\mu}_{\phi}\right],
	\end{eqnarray}
	where we assumed that the functions $f(r)$ and $h(r)=r^2$ transform to $F=F(r,a,\theta)$ and $\Sigma=\Sigma(r,a,\theta)$, respectively. This leads to the following metric
\begin{eqnarray}
ds^2_{4} &=& - F du^2 - 2a\sin^2 \theta [1-F] du d\phi - 2 du dr \nonumber \\\notag
&+& 2a\sin^2 \theta dr d\phi + \Sigma d\theta^2 \\
&+&
[r^2+a^2+a^2\sin^2 \theta \{1-F\}]\sin^2 \theta d\phi^2,
\end{eqnarray}
where $\Sigma \equiv (r^2+a^2\cos^2 \theta)$. Since the geometry in three dimensions can be thought of as the slice of the four dimensional geometry, at this point, we simply set $\theta = \pi/2$ in the metric above to arrive at the rotating black hole metric in three dimensions
\begin{equation}
ds^2 = -F(du - ad\phi)^2 
- 2(du - ad\phi)(dr + ad\phi) + r^2 d\phi^2.
\end{equation} 
It is more suitable to further rewrite this metric in terms of the Boyer-Lindquist-type coordinates
$(t, r, \hat{\phi})$ that is a generalization of the Schwarzschild coordinates. This can be achieved via the transformation
\begin{eqnarray}
dt = du + {(r^2+a^2)\over \Delta}dr, 
~~~d\hat{\phi} = d\phi + {a\over \Delta}dr 
\end{eqnarray}
where $\Delta \equiv r^2 f(r)+a^2$. Finally, the rotating AdS$_{3}$ black hole solution given in Boyer-Lindquist-
type coordinates is given by [henceforth we drop ``hat" on $\phi$ coordinate)]
\begin{eqnarray}
ds^2 &=& -Fdt^2 - 2a[1-F]dt d\phi \\
&+& [r^2+a^2+a^2\{1-F\}]d\phi^2 + {r^2\over \Delta}dr^2. \nonumber
\end{eqnarray}
where there is an expression for $F$ given by [see for details \cite{Azreg-Ainou:2014pra}]
\begin{eqnarray}
F=\frac{\left(f(r)h(r)+a^2\cos^2\theta\right)\Sigma}{\left(k(r)+a^2\cos^2\theta\right)^2}|_{\theta=\pi/2}=f(r),
\end{eqnarray}
where $k(r)=h(r)$. Remarkably, as was shown in \cite{Kim:1998iw}, one can find a coordinate transformation and relate a new time coordinate $\tilde{t}$ to $t$ via
\begin{eqnarray}
\tilde{t} = t - a\phi,
\end{eqnarray} 
and obtain the rotating BTZ geometry
\begin{equation}
ds^2 = -f(r) d\tilde{t}^2 - 2ad\tilde{t}d\phi + r^2d\phi^2 +{r^2\over \Delta}dr^2,
\end{equation}
where $\tilde{t}$ is the BTZ time coordinate. Furthermore we can  also compare it with the general expression of the rotating BTZ black hole in three dimensions given by
\begin{equation}
ds^2 =-N^2(r)d\tilde{t}^2 + \frac{dr^2}{g^2(r)} + R^2(r)[N^{\phi}(r)d\tilde{t} + d\phi]^2.
\end{equation}
Introducing $a=J/2$, it is not difficult to show the following relations:
\begin{eqnarray}
N^2(r)&=&g^2(r)= f(r)+{J^2 \over  4 r^2},\\ \notag
N^{\phi}(r) &=& -{J \over 2 r^2},\\\notag
R^2(r) &=& r^2,\\ \notag
f(r)&=&\frac{r^2}{l^2}-\frac{M r^2}{r^2+l_0^2}-\frac{ Q^2l_0^2}{r^2+l_0^2}- Q^2\ln(r^2+l_0^2).
\end{eqnarray}
Note that, we interpret the above solution only as an effective geometry derived by the dimensional continuation and the NJ algorithm. Nonetheless, it is quite interesting that from the dimensional reduction we can end up with a legitimate rotating black hole solution in three dimensions. \\

\section{Conclusions}
  In this work, we used a regularisation procedure given by the replacement $r\to \sqrt{r^2+l_0^2}$, with $l_0$ being the zero point length and encodes the quantum gravity effects. Then we first modified the gravitational and electromagnetic potentials, we find the energy density of the smeared matter distribution and the energy momentum components of the electromagnetic field. The energy density of the smear matter distribution is regular and finite at the origin while the mass function vanishes at the origin and correctly reduces to the mass parameter for large distances. With these expressions in hand, from the Einstein field equations, we then obtained an exact charged black hole solution in three dimensional AdS space which reduces to the BTZ solution in the limit $l_0 \to 0$. Moreover, we have shown that the black hole solution is regular at the origin, it is asymptotically AdS for large radial coordinate and, interestingly, in specific conditions, due to the quantum gravity effects the solution has a flat de Sitter core at the origin when $r\to 0$.  Finally, using a dimensional continuation, i.e., by assuming that the three black hole geometry is a $\theta = \pi/2$ slice of a static and spherically symmetric four geometry, we applied the complex coordinate transformation via NJ algorithm to obtain the rotating black hole solution to three dimensions. In the near future we plan to explore in more details the thermodynamics, stability, and other properties of the black hole solution reported in this work. 


\end{document}